\documentclass[peerreview, 12pt, draftclsnofoot, onecolumn]{IEEEtran}

\usepackage{cite}
\usepackage{amsmath,amsthm}
\usepackage{array}
\usepackage{amssymb}
\usepackage{array}
\usepackage{cite}
\usepackage{ifthen}
\usepackage{graphicx,subfigure}
\usepackage{psfrag}
\usepackage{dsfont}
\newcommand{\by}{\mathbf{y}}
\newcommand{\bx}{\mathbf{x}}
\newcommand{\ba}{\mathbf{a}}
\newcommand{\ry}{\mathrm{Re}\{\by\}}
\newcommand{\iy}{\mathrm{Im}\{\by\}}
\newcommand{\li}{L^{(i)}}
\newcommand{\lhi}{\hat L^{(i)}}

\newcommand{\lhio}{{\hat L}_0^{(i)}}
\newcommand{\lhii}{{\hat L}_1^{(i)}}
\newcommand{\opt}{\mathrm{opt}}
\newcommand{\Real}{\mathrm{Re}}
\newcommand{\Imag}{\mathrm{Im}}

\newtheorem{thm}{Theorem}

\begin{document}
\title{Efficient LLR Calculation for Non-Binary Modulations over Fading Channels}
\author{Raman~Yazdani,~\IEEEmembership{Student~Member,~IEEE,}
        \\and\\
        Masoud~Ardakani,~\IEEEmembership{Senior~Member,~IEEE}%
        \\
        \normalsize{Department of Electrical and Computer Engineering,
        University of Alberta\\
        Edmonton, Alberta, T6G 2V4, Canada\\
        \{yazdani, ardakani\}@ece.ualberta.ca}}
\maketitle

\begin{abstract}
Log-likelihood ratio (LLR) computation for non-binary modulations
over fading channels is complicated. A measure of LLR accuracy on
asymmetric binary channels is introduced to facilitate good LLR
approximations for non-binary modulations. Considering piecewise
linear LLR approximations, we prove convexity of optimizing the
coefficients according to this measure. For the optimized
approximate LLRs, we report negligible performance losses compared
to true LLRs.
\end{abstract}


\newpage
\section{Introduction}
It is well known that soft-decision decoding algorithms outperform
hard-decision decoding algorithms. In soft-decision decoding,
reliability metrics are calculated at the receiver based on the
channel output. The decoder uses these reliability measures to gain
knowledge of the transmitted codewords. The superiority of soft
decoding comes at the expense of higher complexity.

Log-likelihood ratios (LLRs) have been shown to be very efficient
metrics for soft decoding of many powerful codes such as the
convolutional codes \cite{lin_book}, turbo codes \cite{berrou93},
low-density parity-check (LDPC) codes \cite{gallager63}. LLRs offer
practical advantages such as numerical stability and simplification
of many decoding algorithms. Moreover, due to some properties of the
probability density function (pdf) of the LLRs, such as symmetry and
invariance, LLRs are used as convenient tools for the performance
analysis of binary linear codes
\cite{land00,richardson01design,abedi07}. On many communication
channels, even for binary modulations, i.e., binary phase-shift
keying (BPSK), channel LLRs are complicated functions of the channel
output \cite{hagenauer80}. This fact greatly increases the
complexity of the LLR calculation modules in the decoder causing
decoding delays and power dissipation. In high speed wireless
transmissions, the decoder may not be able to handle this
complexity. Thus, for an efficient implementation of the decoder,
approximate LLRs should be considered.

Approximate LLRs have been previously used in the literature
\cite{hagenauer80,hou01,tosato02,kwon03}. Piecewise linear LLRs have
been suggested in \cite{tosato02} for soft Viterbi decoding of
convolutional codes in the HIPERLAN/2 standard \cite{hiperlan01}.
The presented method uses the log-sum approximation which is quite
accurate at high signal-to-noise ratio (SNR). Moreover, it assumes
that perfect channel state information (CSI) is available at the
receiver. In \cite{yazdani09}, linear LLRs have been used for BPSK
modulation on uncorrelated fading channels without CSI and a measure
of LLR accuracy has been introduced. Using that measure, linear LLR
approximating functions have been designed with almost no
performance gap to that of true LLR calculation. The proposed
measure, however, is only applicable to symmetric channels and BPSK.
With non-binary modulations, used in most practical systems, a
linear approximation of bit LLRs is not always possible. Moreover,
the equivalent bit-channels are asymmetric. That is, the LLR
accuracy measure of \cite{yazdani09} is not applicable.

In this work, we seek approximate LLRs for non-binary signalling
over uncorrelated fading channels. Thus, we need to generalize the
accuracy measure of \cite{yazdani09} to asymmetric channels.
Moreover, we must consider non-linear LLR approximating functions.
To compute LLRs at bit levels individually, we consider
bit-interleaved coded modulation (BICM) \cite{caire98}. BICM is a
well-known bandwidth efficient scheme for binary codes on fading
channels.


While our approaches are general, to demonstrate our methods, we
focus on piecewise linear approximations. Such approximations are
easy to implement and we observe that, when the parameters are
optimized, they perform close to true LLRs. We prove that the
optimization of these piecewise linear approximations is convex,
thus, very efficient. Due to the close-to-capacity performance of
LDPC codes on many channels \cite{chung01,luby01erasure,hou01}, we
employ LDPC-coded BICM \cite{hou03} to show that even with the
proposed approximation, close-to-capacity performance is obtained.


\vspace{-0.3cm}
\section{Problem Definition} \label{sec:prblm_def}
Consider a flat slow-fading environment where the received signal is
expressed as \vspace{-0.2cm}
\begin{equation*}
\mathbf{y}=r\cdot\mathbf{x}+\mathbf{z},
\vspace{-0.2cm}
\end{equation*}
where $\mathbf{x}$ is the complex transmitted signal chosen from the
signal set $\mathcal{X}\subseteq\mathbb{C}$ of size
$|\mathcal{X}|=2^m$, $r\ge0$ is the channel fading gain with
arbitrary pdf $p(r)$, and $\mathbf{z}$ is the additive noise which
is a complex zero-mean white Gaussian random variable with variance
$2\sigma^2$. \vspace{-0.3cm}
\subsection{LLRs for equivalent bit-channels}
Using the BICM scheme \cite{caire98}, the information sequence is
first encoded by a binary code. Next, the coded sequence is bit
interleaved and is broken into $m$-bit sequences which are then Gray
labeled onto signals in $\mathcal{X}$ and transmitted on the
channel. Assuming ideal interleaving, the system can be seen
equivalently as $m$ parallel independent and memoryless binary-input
\emph{bit-channels}. In the receiver, based on $\by$, LLRs are
computed for each bit-channel independently from other bits. These
LLRs are then de-interleaved and passed to the decoder.

When the channel fading gain $r$ is known at the receiver for each
channel use, the true LLR for the $i$th bit-channel, assuming
uniform input distribution, is calculated as
\begin{equation}\label{eq:true_llr_csi}
    \li=\log\frac{P(\mathbf{y}|b^i(\mathbf{x})=0,r)}{P(\mathbf{y}|b^i(\mathbf{x})=1,r)}=\log
    \frac{\sum_{\mathbf{x}\in\mathcal{X}_0^i}p(\mathbf{y}|\mathbf{x},r)}{\sum_{\mathbf{x}\in\mathcal{X}_1^i}p(\mathbf{y}|\mathbf{x},r)}=g_r^{(i)}(\by),
\end{equation}
where $i\in\{1,\dots,m\}$, $b^i(\mathbf{x})$ is the $i$th bit of the
label of $\mathbf{x}$, $\mathcal{X}_w^i$ is the subset of signals in
$\mathcal{X}$ where $b^i(\mathbf{x})=w$, and the conditional
distributions are given by
$p(\mathbf{y}|\mathbf{x},r)=\frac{1}{2\pi\sigma^2}\exp{(-\frac{|\mathbf{y}-r\mathbf{x}|^2}{2\sigma^2})}$
or $\frac{1}{\sqrt{2\pi}\sigma}\exp{(-\frac{(y-rx)^2}{2\sigma^2})}$
when the signal set is real. Also, $g_r^{(i)}(\by)$ represents $\li$
as a function of $\by$ when $r$ is known. When $r$ is not known at
the receiver, the true LLR is calculated as
\begin{equation}\label{eq:true_llr_ncsi}
    \li=\log\frac{P(\mathbf{y}|b^i(\mathbf{x})=0)}{P(\mathbf{y}|b^i(\mathbf{x})=1)}=\log
    \frac{\sum_{\mathbf{x}\in\mathcal{X}_0^i}p(\mathbf{y}|\mathbf{x})}{\sum_{\mathbf{x}\in\mathcal{X}_1^i}p(\mathbf{y}|\mathbf{x})}=g^{(i)}(\by),
\end{equation}
where
$p(\mathbf{y}|\mathbf{x})=\int_0^\infty\frac{1}{2\pi\sigma^2}\exp{(-\frac{|\mathbf{y}-r\mathbf{x}|^2}{2\sigma^2})}p(r)\mathrm{d}r$,
and $g^{(i)}(\by)$ represents $\li$ as a function of $\by$.

As can be seen from (\ref{eq:true_llr_csi}) and
(\ref{eq:true_llr_ncsi}), both $g_r^{(i)}(\by)$ and $g^{(i)}(\by)$
are usually complicated functions of $\by$. Thus, approximate LLRs
($\hat g_r^{(i)}(\by)$ and $\hat g^{(i)}(\by)$) are of practical
interest. One approximation which is useful at high SNR is obtained
by the log-sum approximation: $\log\sum_kz_k\approx\max_k\log z_k$.
This approximation is good when the sum is dominated by a single
large term. Thus,
\begin{equation}\label{eq:logsum}
 \hat g_r^{(i)}(\by)=\log \frac{\max_{\mathbf{x}\in\mathcal{X}_0^i}p(\mathbf{y}|\mathbf{x},r)}
 {\max_{\mathbf{x}\in\mathcal{X}_1^i}p(\mathbf{y}|\mathbf{x},r)},
\end{equation}
\begin{equation*}
\hat g^{(i)}(\by)= \log \frac{\max_{\mathbf{x}\in\mathcal{X}_0^i}p(\mathbf{y}|\mathbf{x})}
{\max_{\mathbf{x}\in\mathcal{X}_1^i}p(\mathbf{y}|\mathbf{x})}=
\log \frac{\max_{\mathbf{x}\in\mathcal{X}_0^i}\int_0^\infty p(\mathbf{y}|\mathbf{x},r)p(r)\mathrm{d}r}
{\max_{\mathbf{x}\in\mathcal{X}_1^i}\int_0^\infty p(\mathbf{y}|\mathbf{x},r)p(r)\mathrm{d}r}.
\end{equation*}
The log-sum approximation is particularly useful when CSI is
available at the receiver, where (\ref{eq:logsum}) leads to
piecewise linear LLRs which can be efficiently implemented
\cite{tosato02}. However, with no CSI, the log-sum approximation is
no longer piecewise linear and in fact involves complicated
integrations. The focus of our work is on cases that CSI is not
available. Nonetheless, we seek approximate LLRs which are piecewise
linear functions of $\by.$

Now, consider a general LLR approximation function parameterized by
set of parameters $\mathcal{A}_i$
\begin{equation*} \label{eq:general_approx__function}
\lhi=\hat g_{\mathcal{A}_i}^{(i)}(\by).
\end{equation*}
This function maps the complex received signal $\by$ to real-valued
approximate LLR for the $i$th bit-channel. Clearly, it is desired to
choose $\mathcal{A}_i$ such that accurate LLR estimates are
obtained. To this end, we first extend the LLR accuracy measure
introduced in \cite{yazdani09} to non-symmetric bit-channels
obtained via BICM. We then use this measure for optimizing the LLR
approximating functions. \vspace{-0.3cm}
\subsection{LLR accuracy measure for asymmetric binary-input channels}

In this section, we generalize the LLR accuracy measure of
\cite{yazdani09} to asymmetric channels. To this end, we consider
the pdfs of the $i$th bit-channel LLR conditioned on the transmitted
bit $b\in\{0,1\},$ defined as
$p_b^{(i)}(l)=E_{\bx\in\mathcal{X}_b^{i}}[p(\li=l|\bx)].$ Using
these LLR pdfs, it is possible to calculate the capacity of each
asymmetric bit-channel and thus the BICM scheme. By capacity, we
mean the mutual information between the input and output of each
bit-channel when its input $b$ is equally likely $0$ or $1$. The
capacity of the $i$th bit-channel is given by \cite{sanaei09}
\begin{equation}\label{eq:cap_BICM_ith}
C_i = 1-\frac{1}{2}\int\log_2(1+e^{-l})p_0^{(i)}(l)\mathrm{d}l
-\frac{1}{2}\int\log_2(1+e^{l})p_1^{(i)}(l)\mathrm{d}l.
\end{equation}
Thus, the capacity of the BICM is found by
$C=\sum_{i=1}^mC_i.$

When instead of the true LLRs, approximate LLRs $\lhi$ are used,
their pdfs are given by $\hat
p_b^{(i)}(l)=E_{\bx\in\mathcal{X}_b^{i}}[p(\lhi=l|\bx)]$ for
$b\in\{0,1\}$. Inserting these approximate LLR pdfs in
(\ref{eq:cap_BICM_ith}) we get
\begin{equation}\label{eq:cap_BICM_approx}
\hat C = \sum_{i=1}^m\hat C_i = \sum_{i=1}^m\left(1-\frac{1}{2}\int\log_2(1+e^{-l})\hat p_0^{(i)}(l)\mathrm{d}l
-\frac{1}{2}\int\log_2(1+e^{l})\hat p_1^{(i)}(l)\mathrm{d}l\right).
\end{equation}
The following theorem, proved in Appendix A, shows $\hat C$ can be
used as an LLR accuracy measure.
\begin{thm}\label{th:measure}
The maximum of $\hat C$ in (\ref{eq:cap_BICM_approx}) is equal to
$C$ which is achieved by true LLRs (no LLR approximation can result
in $\hat C>C$).
\end{thm}

Notice that for a symmetric channel (i.e.,
$p_1^{(i)}(l)=e^{-l}p_0^{(i)}(l)$), (\ref{eq:cap_BICM_approx})
reduces to the LLR accuracy measure of \cite{yazdani09}. Using
similar arguments, it can be shown that as the approximate LLRs
drift away more from the true LLRs, $\Delta C=C-\hat C$ gets larger.
Thus, $\hat C$ is a measure of the accuracy of the approximate bit
LLRs. Since bit-channels are independent, we maximize each $\hat
C_i$ individually. Thus, for each bit-channel $i$, assuming a class
of approximating functions $\hat g_{\mathcal{A}_i}^{(i)}(\by)$, we
find:
\begin{eqnarray}\label{eq:optimization}
    \mathcal{A}_i^{\opt}&=&\arg\max_{\mathcal{A}_i}\hat C_i,\\
    \nonumber&&\mathrm{s.t.}\;
    \,{\scriptstyle\mathbf{\Phi}_i(\mathcal{A}_i)=0}
\end{eqnarray}
where $\mathbf{\Phi}_i(\mathcal{A}_i)=0$ denotes the constraints
imposed on $\mathcal{A}_i$ (e.g., to preserve continuity).
\vspace{-0.3cm}
\subsection{LLR approximating functions}
It is evident from (\ref{eq:true_llr_csi}) and
(\ref{eq:true_llr_ncsi}) that true LLR functions depend on the
signal set $\mathcal{X}$. Thus, choosing appropriate class of
approximating functions also depends on $\mathcal{X}$. In this
letter, we consider non-binary AM and rectangular QAM. Our
optimization approach, however, is general. Moreover, we put our
focus on piecewise linear approximate LLRs. Clearly, the
optimization problem (\ref{eq:optimization}) can be solved for any
other approximating function. Using piecewise linear approximations
has benefits such as simplicity of demodulator and (as will be
shown) convexity of the optimization problem. Numerical results
verify that the obtained performance is also very close to true
LLRs.



By viewing a complex variable as a two-dimensional vector
$\by=(\Real\{\by\},\Imag\{\by\})$, a piecewise linear function of a
complex variable is defined as follows. First, the complex domain
$\mathbb{C}$ is divided into a finite number of regions
$\mathbb{C}_1,\mathbb{C}_2,\dots,\mathbb{C}_N$ by a finite number of
one-dimensional boundaries. Then the function is represented by
$f(\by)=\langle\mathbf{a}_k,\by\rangle+b_k$ for any
$\by\in\mathbb{C}_{k}$, where $\langle,\rangle$ denotes the inner
product of two-dimensional vectors, i.e.,
$\langle\mathbf{a}_k,\by\rangle=\Real\{\ba_k\}\Real\{\by\}+\Imag\{\ba_k\}\Imag\{\by\}$.
Thus,
\begin{eqnarray}\label{eq:piecewise_linear}
\lhi=\hat g_{\mathcal{A}_i}^{(i)}(\by)&=&\sum_{k=1}^{N^{(i)}}\left(\langle\ba_k^{(i)},\by\rangle+b_k^{(i)}\right)\mathbf{1}_{(\by\in \mathbb{C}_k^{(i)})},
\end{eqnarray}
where $N^{(i)}$ is the number of segments of the piecewise linear
function, $\mathcal{A}_i$ is the set of all $\ba_k^{(i)}$,
$b_k^{(i)}$, and $\mathbf{1}_{(\cdot)}$ is the indicator function.
The parameters are chosen to preserve continuity over $\by$.

The following theorem, proved in Appendix B, indicates that
optimizing a piecewise linear approximating function according to
(\ref{eq:optimization}) is a convex optimization problem.
\begin{thm} \label{th:convex}
Assuming that approximate bit LLRs are calculated by
(\ref{eq:piecewise_linear}) for $i=1,\dots,m$, and assuming fixed
$\mathbb{C}_k^{(i)}$ for $k=1,\dots,N^{(i)}$, $\hat C_i$ is a
concave function of $\ba_k^{(i)}$ and $b_k^{(i)}$ for all
$k.$
\end{thm}

Using (\ref{eq:piecewise_linear}), (\ref{eq:optimization}) can be
numerically solved as follows. For a given SNR, and assuming fixed
$\mathbb{C}_k^{(i)}$'s, $\hat C_i$ can be computed by first
computing $\hat p_0^{(i)}(l)$ and $\hat p_1^{(i)}(l)$ for given
$\ba_k^{(i)}$'s and $b_k^{(i)}$'s and inserting them in
(\ref{eq:cap_BICM_approx}). Since $\hat C_i$ is a concave function
of $\ba_k^{(i)}$'s and $b_k^{(i)}$'s and the constraints are linear,
maximizing $\hat C_i$ can be done efficiently using numerical
optimization techniques. The proper number of regions $N^{(i)}$ is
based on the affordable complexity and the curve of true LLRs.
Optimizing the regions can be done through search. As will be seen
in the next section, usually the size of the parameter sets is small
and due to the symmetry in the LLR calculation, many parameters are
equal to each other which can further reduce the number of unknown
parameters.

\vspace{-0.3cm}
\section{Proposed Approach and Examples}\label{sec:approach}
Now, we describe the proposed method through examples of real and
complex signal constellations.

\begin{figure}[p]
\centering
\includegraphics[width=0.85\columnwidth]{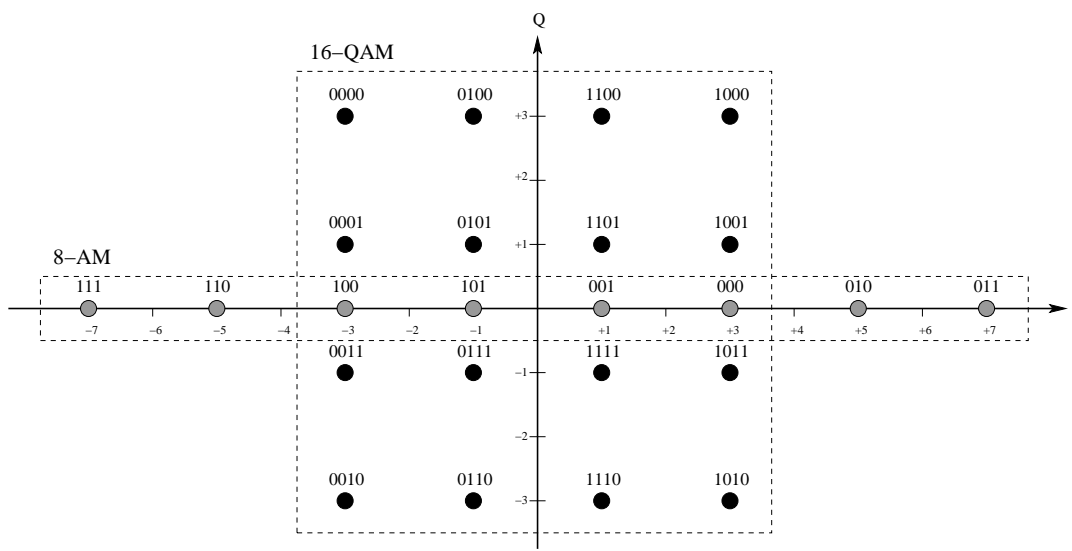}\\
\caption{The 8-AM and 16-QAM constellation points and Gray mapping.}\label{fig:constellations}
\end{figure}

\textbf{Example~1:} Consider 8-AM constellation with Gray labeling
shown in Fig.~\ref{fig:constellations} on the normalized Rician
fading channel. On this channel,
$p(r)=2r(K+1)e^{-(K+(K+1)r^2)}I_0(2r\sqrt{K(K+1)})$, where $K$ is
the Rician K-factor, and $I_0(\cdot)$ is the zero-order modified
Bessel function of the first kind.

\begin{figure}[!t]
\centering
    \subfigure[$i=1$]
    {
      \label{fig:8_PAM_real1}{\includegraphics[width=.48\columnwidth]{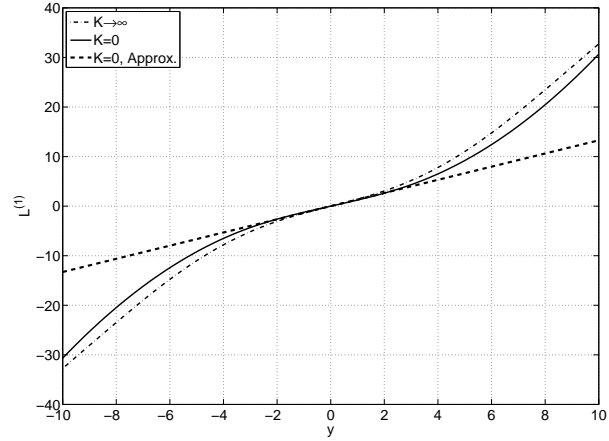}}
    }
    \hfill
    \subfigure[$i=2$]
    {
      \label{fig:8_PAM_real2}{\includegraphics[width=.48\columnwidth]{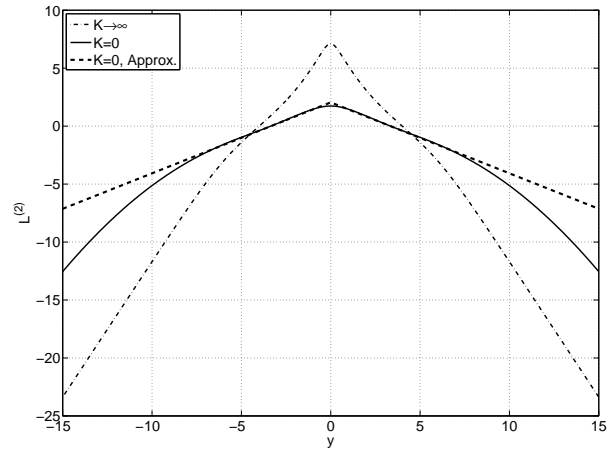}}
    }
    \hfill
    \subfigure[$i=3$]
    {
      \label{fig:8_PAM_real3}{\includegraphics[width=.48\columnwidth]{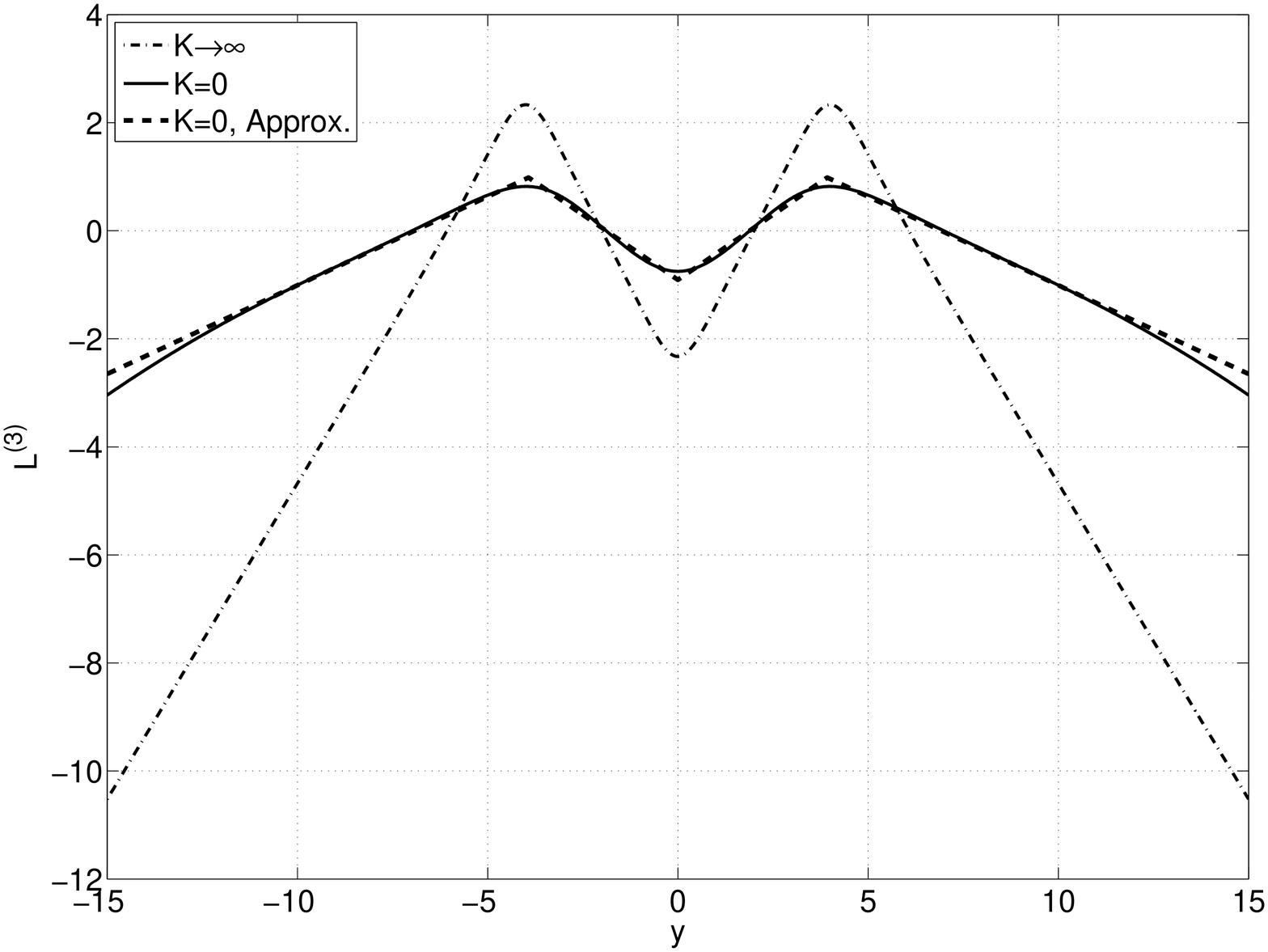}}
    }
  \caption{True bit LLR values $\li$ as functions of the channel output $y$ for the 8-AM at SNR$=7.88$~dB. Also, the piecewise linear LLR
  approximations with optimized parameters of Table~\ref{tb:opt_param_(3,4)} are depicted for $K=0$. }
  \label{fig:8_PAM_real}
\end{figure}

Using (\ref{eq:true_llr_csi}) and (\ref{eq:true_llr_ncsi}), true
LLRs are calculated for $i=1,2,3$, and they have been plotted versus
$y$ for extreme values of $K$ in Fig.~\ref{fig:8_PAM_real}.
Considering the general model of (\ref{eq:piecewise_linear}), and
the symmetry in the true LLR functions, we propose the following
piecewise linear LLR approximations
\begin{eqnarray}
\hat {L}^{(1)}=\hat g_{\mathcal{A}_1}^{(1)}(y)&=&a_1^{(1)} y, \label{eq:8PAM_linear_1}\\
\hat {L}^{(2)}=\hat g_{\mathcal{A}_2}^{(2)}(y)&=&(a_1^{(2)} y+b_1^{(2)})\mathbf{1}_{(y\leq0)}+(a_2^{(2)} y+b_2^{(2)})\mathbf{1}_{(0<y)}=-a_1^{(2)}|y|+b_1^{(2)},\label{eq:8PAM_linear_2}\\
\nonumber\hat {L}^{(3)}=\hat g_{\mathcal{A}_3}^{(3)}(y)&=&(a_1^{(3)} y+b_1^{(3)})\mathbf{1}_{(y\le c_1^{(3)})}+(a_2^{(3)}y+b_2^{(3)})\mathbf{1}_{(c_2^{(3)}< y\le0)}\\
&& +(a_3^{(3)} y+b_3^{(3)})\mathbf{1}_{(0<y\le c_3^{(3)})}+(a_4^{(3)} y+b_4^{(3)})\mathbf{1}_{(c_4^{(3)}<y)}\label{eq:8PAM_linear_3},
\end{eqnarray}
where due to the symmetry of the LLRs, we have assumed in
(\ref{eq:8PAM_linear_2}) that $a_2^{(2)}=-a_1^{(2)}$ and
$b_2^{(2)}=b_1^{(2)}$. Also, in (\ref{eq:8PAM_linear_3}), we have
$a_1^{(3)}=-a_4^{(3)}$, $a_2^{(3)}=-a_3^{(3)}$,
$b_1^{(3)}=b_4^{(3)}$, $b_2^{(3)}=b_3^{(3)}$, and
$c_1^{(3)}=c_2^{(3)}=-c_3^{(3)}=-c_4^{(3)}$. Thus,
$\mathcal{A}_1=\{a_1^{(1)}\}$,
$\mathcal{A}_2=\{a_1^{(2)},b_1^{(2)}\}$, and
$\mathcal{A}_3=\{a_1^{(3)},a_2^{(3)},b_1^{(3)},b_2^{(3)},c_1^{(3)}\}$.
It is evident that symmetry reduces the number of unknown
parameters. We also impose
$\Phi(\mathcal{A}_3)=c_1^{(3)}(a_1^{(3)}-a_2^{(3)})+b_1^{(3)}-b_2^{(3)}=0$
to preserve continuity in (\ref{eq:8PAM_linear_3}).
Fig.~\ref{fig:8_PAM_real} shows that these piecewise linear
functions better approximate the true LLRs when $K$ increases.

For a given SNR, we optimize the parameter sets $\mathcal{A}_1$,
$\mathcal{A}_2$, and $\mathcal{A}_3$, by solving
(\ref{eq:optimization}). Numerical results confirm that $\hat
C_{\mathrm{max}}=\sum_{i=1}^m\max_{\mathcal{A}_i}\hat C_i$ is always
very close to the capacity of BICM employing true LLRs, i.e., $C$.
For example, for $K=0$, $\hat C_{\mathrm{max}}=0.851$ and $C=0.855$
bits per channel use at SNR$=5.00$~dB, and $\hat
C_{\mathrm{max}}=1.544$ and $C=1.553$ bits per channel use at
SNR$=30.00$~dB. When $K$ increases, $\Delta C=C-\hat
C_{\mathrm{max}}$ becomes even smaller.



To evaluate the decoding performance of the optimized piecewise
linear approximations, we compare the decoding threshold of LDPC
codes and their bit error rate (BER) under approximate and true
LLRs. The decoding threshold can be found by density evolution
\cite{richardson01design,chung01} and by using the technique of
i.i.d. channel adapters \cite{hou03} which provides the required
symmetry conditions.

As an example, consider $(3,4)$-regular LDPC codes on the normalized
Rayleigh channel (equivalent to $K=0$) with 8-AM signalling of
Fig.~\ref{fig:constellations}. By using the approximating functions
of (\ref{eq:8PAM_linear_1})--(\ref{eq:8PAM_linear_3}), we find the
decoding threshold of the code under optimized LLR parameters
reported in Table~\ref{tb:opt_param_(3,4)}. The decoding threshold
given by density evolution is 7.88~dB while under true LLR
calculation of (\ref{eq:true_llr_ncsi}), the decoding threshold is
7.85~dB showing only a 0.03~dB performance gap.

\begin{figure}
\centering
\includegraphics[width=0.8\columnwidth]{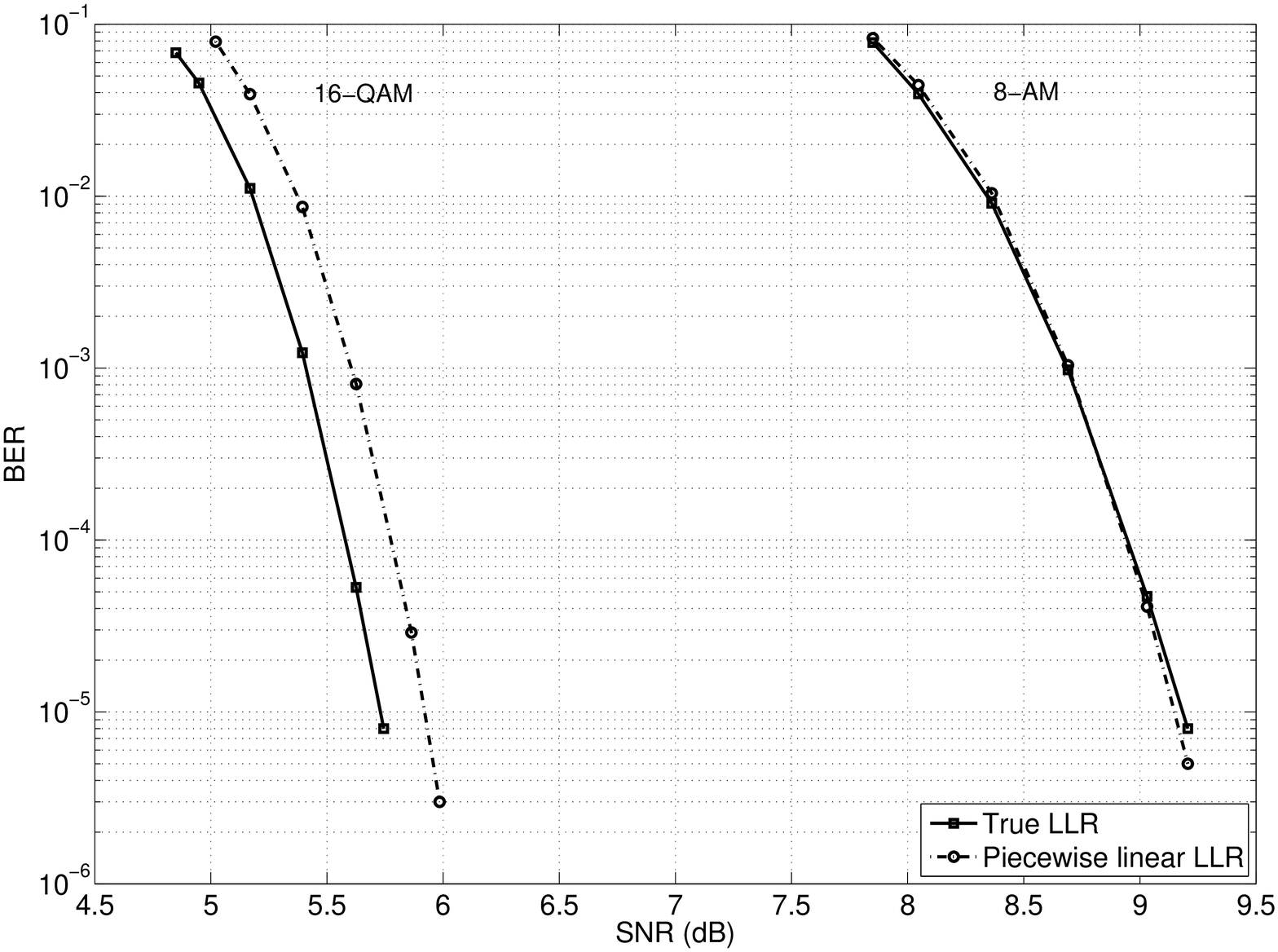}\\
\caption{Comparison between the BER of a randomly constructed $(3,4)$-regular LDPC code of length $15000$ decoded
by true and approximate LLRs on the Rayleigh fading channel ($K=0$). The approximate LLR parameters are reported in Table~\ref{tb:opt_param_(3,4)}.}\label{fig:BER_3_4}
\end{figure}

To see how the piecewise linear LLR calculation affects the BER
performance, we simulate a given LDPC code on the normalized
Rayleigh fading channel. In Fig.~\ref{fig:BER_3_4}, the performance
of a randomly constructed $(3,4)$-regular LDPC code of length
$15000$ is depicted in two cases: once decoded using true LLRs of
(\ref{eq:true_llr_ncsi}), and once with the piecewise linear
approximation of (\ref{eq:8PAM_linear_1})--(\ref{eq:8PAM_linear_3})
and the optimized parameters reported in
Table~\ref{tb:opt_param_(3,4)}. It should be noted that the
parameters are optimized once at the decoding threshold and are kept
fixed at the receiver for other SNRs. It is seen that the
performance of the optimized approximate LLRs is almost identical to
that of the more complex true LLRs although parameters are only
optimized for the worst SNR.

Also, to show that optimizing $\hat C$ is meaningful in terms of the
maximum transmission rate achievable by the piecewise linear LLRs,
we optimize the degree distributions \cite{luby01} of LDPC codes
under our approximate LLRs. At SNR $=21.02$~dB, the capacity of BICM
under true LLRs is $C=1.500$ in the absence of CSI when $K=0$. Since
$m=3$, then the maximum binary code rate achievable on this channel
is $0.500$. At this SNR, solving (\ref{eq:optimization}) gives $\hat
C_{\mathrm{max}}=1.493$ and the parameters reported in
Table~\ref{tb:opt_param_(3,4)}. Now, by using the designed piecewise
linear approximation, assuming a fixed check node degree of $8$ and
maximum variable node degree of $30$, an irregular LDPC code is
designed. The variable node degree distribution of this code is
$\lambda(x)=0.250x\!+\!0.217x^2\!+\!0.221x^6\!+\!0.048x^7\!+\!0.119x^{22}\!+\!0.145x^{29}$,
and the code rate is $R=0.490$. Thus, the proposed approximate LLRs
can achieve rates very close to the capacity of BICM under true
LLRs.

\textbf{Example~2:} Now consider a 16-QAM constellation with Gray
labeling as depicted in Fig.~\ref{fig:constellations}. Using the
general piecewise linear model of (\ref{eq:piecewise_linear}), due
to the symmetry and the similarity of the bit LLR functions, we
propose the following LLR approximations:
\begin{eqnarray}
\hat {L}^{(1)}=&\hat g_{\mathcal{A}_1}^{(1)}(\by)&=a_1^{(1)} \ry, \label{eq:16QAM_linear_1}\\
\nonumber\hat {L}^{(2)}=&\hat g_{\mathcal{A}_2}^{(2)}(\by)&=\sum_{k=1}^4\left(\langle\ba_k^{(2)},\by\rangle+b_k^{(2)}\right)\mathbf{1}_{(\by\in\mathbb{C}_k^{(2)})}\\
&&=\mathrm{Re}\{\ba_1^{(2)}\} |\ry|+\mathrm{Im}\{\ba_1^{(2)}\}|\iy|+b_1^{(2)},\label{eq:16QAM_linear_2}
\end{eqnarray}
where $\mathbb{C}_1^{(2)},\dots,\mathbb{C}_4^{(2)}$ are the four
quadrants of the complex plane. It should be noted that bit-channel
LLR calculations are similar for bit 1 and 3, and for bit 2 and 4
except that the real and imaginary parts of $\by$ are swapped, i.e.,
$\hat {L}^{(3)}=a_1^{(1)} \iy$ and $\hat
{L}^{(4)}=\mathrm{Re}\{\ba_1^{(2)}\}
|\iy|+\mathrm{Im}\{\ba_1^{(2)}\}|\ry|+b_1^{(2)}$. Thus, it is enough
to optimize $\mathcal{A}_1=\{a_1^{(1)}\}$ and
$\mathcal{A}_2=\{\ba_1^{(2)},b_1^{(2)}\}$.

Again numerical results suggest that the gap between $\hat
C_{\mathrm{max}}$ and the true BICM capacity is always small. For
example, when $K=0$, we have $\hat C_{\mathrm{max}}=1.074$ and
$C=1.097$ bits per channel use at SNR=$3.00$~dB. Again, the gap
becomes smaller when $K$ increases.



To investigate the performance of 16-QAM signalling under the
proposed approximate LLRs, we compare the decoding threshold and BER
of $(3,4)$-regular LDPC codes on the Rayleigh fading channel under
true and approximate LLRs. Density evolution gives a decoding
threshold of $5.02$~dB under approximate LLRs with the optimized
parameters of Table~\ref{tb:opt_param_(3,4)}. Under true LLR
calculation, the decoding threshold is $4.83$~dB. As a result,
approximate LLRs show about $0.19$~dB performance gap to true LLRs.
The BER comparison is depicted in Fig.~\ref{fig:BER_3_4}.  It is
worth mentioning that this gap can be further reduced by proposing
piecewise linear LLRs with more segments.

\vspace{-0.3cm}
\section{Conclusion} \label{sec:conclusion}
LLR computation for equivalent bit-channels of a non-binary
modulation is generally complicated. On fading channels, when the
channel gain is unknown, this problem is further intensified.
Noticing that the equivalent bit channels were asymmetric, in order
to find good approximate LLRs, we proposed an LLR accuracy measure
for binary asymmetric channels. This accuracy measure can be used to
optimize the parameters of any approximating function. We used our
accuracy measure to optimize piecewise linear LLR approximations. By
using LDPC-coded BICM, we showed that the performance loss under the
optimized piecewise linear approximation was very small.  We also
showed that under approximate LLRs, asymptotic irregular LDPC codes
having rates very close to the capacity of BICM under true LLRs can
be obtained. Our solution can also be applied to other coding
schemes which use LLRs such as the convolutional and turbo codes.
\vspace{-0.4cm} \appendices

\section{Proof of Theorem~\ref{th:measure}}\label{App:3}
Consider an arbitrary discrete binary-input memoryless channel whose
output alphabet is non-binary. The channel input is $x \in \{0,1\}$,
and its output is $y \in \{ y_j| 1\le j \le M \}$. Let us define
$P(y_j|x=0)=p_j$ and $P(y_j|x=1)=q_j$ where
$\sum_{j=1}^{M}{p_j}=\sum_{j=1}^{M}{q_j}=1$. The true LLR value,
when $y=y_j$ is observed at the channel output and the binary inputs
are equiprobable, is
\begin{equation}\label{eq:app_true_llr}
l_j=g(y_j)=\log\frac{p_j}{q_j}.
\end{equation}
Thus, the true LLR pdf when $x=0$ is sent is given by $f_0(l) =
\sum_{j=1}^Mp_j\delta\left(l-\log\frac{p_j}{q_j}\right)$ and by
$f_1(l) = \sum_{j=1}^Mq_j\delta\left(l-\log\frac{p_j}{q_j}\right)$
when $x=1$ is sent over the channel.

Now, assuming that when $y_j$ is observed at the channel output, the
approximate LLR is calculated by $\hat l_j=\hat g(y_j)=a_j$, the
conditional pdfs of $\hat l$ are: \vspace{-0.2cm}
\begin{eqnarray}
f_0(\hat l) &=& \sum_{j=1}^Mp_j\delta\left(\hat l-a_j\right),\label{eq:app_pdf0}\\
f_1(\hat l) &=& \sum_{j=1}^Mq_j\delta\left(\hat
l-a_j\right).\label{eq:app_pdf1}
\end{eqnarray}\vspace{-0.2cm}
Inserting (\ref{eq:app_pdf0}) and (\ref{eq:app_pdf1}) in
(\ref{eq:cap_BICM_ith}) gives
\begin{equation}
\hat C_i=1-\frac{1}{2}\sum_{j=1}^M\left(p_j\log_2(1+e^{-a_j})+q_j\log_2(1+e^{a_j})\right).
\end{equation}
Taking $\frac{\partial\hat C_i}{\partial a_j}$ reveals that
$a_j=\log\frac{p_j}{q_j}$ maximizes $\hat C_i$ for all $1\le j \le
M$ since $\frac{\partial^2\hat C_i}{\partial a_j^2}<0$ for all $1\le
j\le M$ and $\frac{\partial^2\hat C_i}{\partial a_j\partial a_k}=0$
for all $1\le j \le M$ and $1\le k \le M$ and $j\neq k$. These
values of $a_j$'s are equal to the true LLR of
(\ref{eq:app_true_llr}). Thus, the maximizing point is only given by
true LLRs. Noticing that these results are valid for each equivalent
bit-channel $i$ of the BICM and since $\max \hat
C=\sum_{i=1}^m\max_{\mathcal{A}_i}\hat C_i$ in
(\ref{eq:cap_BICM_approx}), the theorem is proved.

\vspace{-0.4cm}
\section{Proof of Theorem~\ref{th:convex}}\label{App:2}
Denote ${\hat L}_b^{(i)}=E_{\bx\in\mathcal{X}_b^{i}}[\lhi|\bx]$ and
$\by_b^{(i)}=E_{\bx\in\mathcal{X}_b^{i}}[\by|\bx]$ for
$b\in\{0,1\}$. Then $\hat C_i$ can be written as
\begin{align*}
\hat C_i=&1-\frac{1}{2}E_{\lhio}\left[\log_2(1+e^{-\lhio})\right]
-\frac{1}{2}E_{\lhii}\left[\log_2(1+e^{\lhii})\right]\\
=&1-\frac{1}{2}E_{\by_0^{(i)}}\left[\log_2\left(1+e^{-\hat g_{\mathcal{A}_i}^{(i)}(\by_0^{(i)})}\right)\right]
-\frac{1}{2}E_{\by_1^{(i)}}\left[\log_2\left(1+e^{\hat g_{\mathcal{A}_i}^{(i)}(\by_1^{(i)})}\right)\right].
\end{align*}
By using (\ref{eq:piecewise_linear}) and with some abuse of notation
we write
\begin{align*}
\hat C_i= 1-\frac{1}{2}\sum_{b=0}^1\sum_{k=1}^{N^{(i)}} E_{(\by_b\in\mathbb{C}_k^{(i)})}\left[\log_2\left(1+e^{(-1)^{b+1}
(\langle\ba_k^{(i)},\by_b\rangle+b_k^{(i)})}\right)\right].
\end{align*}
For each fixed $\by_b\in\mathbb{C}_k^{(i)}$, it is clear that $\hat
g$ is a linear function of $\ba_k^{(i)}$ and $b_k^{(i)}$. Noticing
that the function $\log_2\left(1+\exp(\cdot)\right)$ is convex and
twice differentiable, it can be deduced that
$\log_2(1+\exp((-1)^{b+1}(\langle\ba_k^{(i)},\by_b\rangle+b_k^{(i)})))$
is also a convex function of $\ba_k^{(i)}$ and $b_k^{(i)}$. The
convexity is also preserved under expectation. Thus,
$E_{(\by_b\in\mathbb{C}_k^{(i)})}[\log_2(1+\exp{((-1)^{b+1}(\langle\ba_k^{(i)},\by_b\rangle+b_k^{(i)}))})]$
is also convex which makes $\hat C_i$ concave with respect to
$\ba_k^{(i)}$ and $b_k^{(i)}$ for all $k=1,\dots,N^{(i)}$.
\vspace{-0.3cm}
\bibliographystyle{IEEEtran}

\bibliography{IEEEabrv,mybib}

\begin{table}[p]
\begin{center}
\scriptsize
\begin{tabular}{|c|c|c|c|c|c|c|c|c|}
\hline \multicolumn{9}{|c|}{8-AM}\\
\hline SNR & Bit 1 & \multicolumn{2}{c|}{Bit 2} & \multicolumn{5}{c|}{Bit 3}\\
\hline
7.88~dB & $a_1^{(1)}=1.328$ & $a_1^{(2)}=0.612$ & $b_1^{(2)}=2.046$ & $a_1^{(3)}=0.328$ & $b_1^{(3)}=2.273$ &
$a_2^{(3)}=-0.482$ & $b_2^{(3)}=-0.909$ & $c_1^{(3)}=-3.928$ \\
\hline
21.03~dB & $a_1^{(1)}=8.538$ & $a_1^{(2)}=0.825$ & $b_1^{(2)}=3.098$ & $a_1^{(3)}=0.384$ & $b_1^{(3)}=2.513$ &
$a_2^{(3)}=-1.528$ & $b_2^{(3)}=-2.357$ & $c_1^{(3)}=-2.547$ \\
\hline
\hline \multicolumn{9}{|c|}{16-QAM}\\
\hline \multicolumn{2}{|c|}{SNR} & \multicolumn{2}{c|}{Bit 1} & \multicolumn{5}{c|}{Bit 2}\\
\hline
\multicolumn{2}{|c|}{5.02~dB} & \multicolumn{2}{c|}{$a_1^{(1)}=1.262$} & \multicolumn{3}{c|}{$\ba_1^{(2)}=(0.868,-0.200)$} & \multicolumn{2}{c|}{$b_1^{(2)}=-1.257$}\\
\hline
\end{tabular}
\caption{{Optimized piecewise linear LLR parameters at different
SNRs for 8-AM and 16-QAM when $K=0$.}} \label{tb:opt_param_(3,4)}
\end{center}
\end{table}

\end{document}